\documentclass{emulateapj}
 \usepackage{url}

\shorttitle{Braking Index and Energy Partition of Magnetar spindown}
\shortauthors{L\"{u} et al.}
\slugcomment{}

\begin{document}

\title{Constraining the Braking Index and Energy Partition of Magnetar spindown with {\em Swift}/XRT data}
\author{Hou-Jun L\"{u}, Lin Lan, and En-Wei Liang}
\affil{Guangxi Key Laboratory for Relativistic Astrophysics, School of Physical Science and Technology,
Guangxi University, Nanning 530004, China; lhj@gxu.edu.cn}

\begin{abstract}
The long-lasting X-ray plateau emission in long gamma-ray bursts (GRBs) shows observational evidence
for ongoing energy injection, which may be from magnetar spindown due to energy released via either
magnetic dipole (MD) or gravitational wave (GW) radiation. In this paper, by systematically analyzing
the {\em Swift}/XRT light curves detected before 2018 July, we find 45 light curves with a measured
redshift that monotonically decay as a smooth broken power law. By assuming that the central engines of
these GRBs are newly born magnetars, we measure the braking index $n$ of putative millisecond
magnetars, due to MD and GW radiations. The inferred braking indices are not close to 3 or 5, but range
between them with a normal distribution ($n_{\rm c}=4.02\pm 0.11$). We define a dimensionless parameter
$\Re$, which is the ratio between the MD and GW components, and find that the energy released via
magnetar spindown in most GRBs of our sample is dominated by GW radiation for $P_0=3$ ms and
$\epsilon=0.005$ and 0.01. On the other hand, we find that $\Re$ and the braking index $n$ seem to be
anticorrelated within a large systematic error at $t=0$, but depend on the values of the parameters
$P_0$ and $\epsilon$. These results suggest that the contribution of GW radiation cannot be ignored,
and that a larger braking index leads to GWs dominating the energy released during magnetar spindown if
indeed magnetars are operating in some long GRBs.

\end{abstract}


\section{Introduction}
Millisecond magnetars (namely a rapidly spinning, strongly magnetized neutron stars) are potential
candidates for the central engine of gamma-ray bursts (GRBs), which are thought to be from a violent
event such as the collapse of a massive star (long GRBs) or the coalescence of two compact objects
(short GRBs; Paczynski 1986; Eichler et al. 1989; Usov 1992; Woosley 1993; Thompson 1994; Dai \& Lu
1998a,b; MacFadyen \& Woosley 1999; Zhang \& M\'esz\'aros 2001; Metzger et al. 2008; Zhang 2011). Both
gravitational wave (GW) and magnetic dipole (MD) radiations can be generated by the rotating neutron
star with an asymmetrical mass arrangement (Zhang \& M\'esz\'aros 2001; Yu et al. 2010; Metzger et al.
2011; Fan et al. 2013; Lasky et al. 2014; Lasky \& Glampedakis 2016; L\"{u} et al. 2018). The
long-lasting X-ray plateau emission in both long and short GRBs show observational evidence of ongoing
energy injection from magnetar spindown (Fan \& Xu 2006; Zhang et al. 2006; Liang et al. 2007; Troja et
al. 2007; Lyons et al. 2010; Rowlinson et al. 2010; Bucciantini et al. 2012; Rowlinson et al. 2013;
Gompertz et al. 2013; L\"{u} \& Zhang 2014; L\"{u} et al. 2015; Gao, Zhang \& L\"{u} 2016; Chen et al.
2017; Gibson et al. 2017). Unfortunately, the GW signal produced via a rotating neutron star is too
weak to be detected by the present-day Advanced LIGO and Advanced Virgo (Abbott et al. 2017; L\"{u} et
al. 2017; Sarin et al. 2018).

The energy reservoir of a millisecond magnetar is the total rotation energy, which reads as
\begin{eqnarray}
E_{\rm rot} = \frac{1}{2} I \Omega^{2}
\simeq 2 \times 10^{52}~{\rm erg}~
M_{1.4} R_6^2 P_{-3}^{-2},
\label{Erot}
\end{eqnarray}
where $I$ is the moment of inertia; and $\Omega$, $P$, $R$, and $M$ are the angular frequency, rotating
period, radius, and mass of the neutron star, respectively. The convention $Q = 10^x Q_x$ in cgs units
is adopted. A magnetar spinning down loses its rotational energy via MD torques ($L_{\rm EM}$) and GW
($L_{\rm GW}$) radiation emissions (Zhang \& M{\'e}sz{\'a}ros 2001; Fan et al. 2013; Giacomazzo \&
Perna 2013; Lasky \& Glampedakis 2016),
\begin{eqnarray}
-\frac{dE_{\rm rot}}{dt} = -I\Omega \dot{\Omega} &=& L_{\rm total} = L_{\rm EM} + L_{\rm GW} \nonumber \\
&=& \frac{B^2_{\rm p}R^{6}\Omega^{4}}{6c^{3}}+\frac{32GI^{2}\epsilon^{2}\Omega^{6}}{5c^{5}},
\label{Spindown}
\end{eqnarray}
where $B_{\rm p}$ is the surface magnetic field at the pole and
$\epsilon=2(I_{xx}-I_{yy})/(I_{xx}+I_{yy})$ is the ellipticity describing how large of the neutron star
deformation. $\dot{\Omega}$ is the time derivative of the angular frequency, and it can be described
with the torque equation (Lasky et al. 2017),
\begin{eqnarray}
\dot{\Omega}=-k\Omega^{n},
\label{torque equation}
\end{eqnarray}
where $k$ is a constant of proportionality, and $n$ is the braking index\footnote{Here, we assume that
the value of $k$ is constant, even though it sometimes evolves with time and depends on the equation of
state of the neutron star (Woan et al. 2018).}.

Traditionally, the physical parameters of magnetars in a GRB study have been estimated by fitting the
observed X-ray plateau emission with a magnetar model assuming $n=3$ and ignoring the contribution of
GW radiation (Troja et al. 2007; Lyons et al. 2010; Rowlinson et al. 2010;L\"{u} \& Zhang 2014; L\"{u}
et al. 2015). In principle, the contribution of GW radiation to a rotating neutron star should be
considered, and the braking index should not be a constant (Fan et al. 2013; Lasky \& Glampedakis
2016). If this were the case, the accurate estimated value of the braking index becomes very important
in understanding the properties of the magnetar and the fraction of energy released as GW and MD
radiations (Woan et al. 2018). Lasky et al. (2017) first proposed constraining the braking indices of
magnetars by invoking the X-ray plateau emission of two short GRBs with measured redshifts. However,
they used the observed X-ray plateau luminosity, which is equal to the total luminosity ($L_{\rm
total}$) of the released magnetar energy, to fit the data. In fact, the intrinsic luminosity of GW
radiation should not be accounted for when fitting the observed data, although it can affect
electromagnetic (EM) emission when GW radiation is dominate in or contributes to the spindown of
magnetars. In other words, the $\Omega(t)$ does not evolve with time, neither following $\sim t^{-1/2}$
(MD-dominated) nor $\sim t^{-1/4}$ (GW-dominated) when the time is much longer than the characteristic
spindown timescale (L\"{u} et al. 2018).

In this paper, by considering the contribution of GW radiation to magnetar spindown, we perform a
systematic study of long GRBs, in whose central engine may resid a millisecond magnetar, and
investigate constraining the braking index using the X-ray light curves exhibited by the plateau
following normal decay emissions of long GRBs. We also try to determine the fraction of energy released
by the magnetar as MD and GW radiations if the magnetar is indeed the central engine of those long
GRBs.

This paper is organized as follows. Our sample selection and data for the magnetar model fitting are
presented in section 2. The result of braking index constraint of magnetar is shown in section 2. In
section 3, we roughly calculate the fraction of magnetar energy released as MD and GW radiations. The
conclusions, along with some discussions, are presented in Section 4. Throughout this paper, a
concordance cosmology with parameters $H_{\rm 0} = 71\rm~km~s^{-1}~Mpc^{-1}$, $\Omega_M=0.30$, and
$\Omega_{\Lambda}=0.70$ is adopted.

\section{Sample Selection and Data Fitting with the Magnetar Model}
\subsection{Sample Selection}
Our entire sample includes more than 1300 GRBs observed between 2005 January and 2018 July, the X-Ray
Telescope (XRT) data of which were selected and downloaded from the {\em Swift} data archive and the UK
{\em Swift} Science Data Center\footnote{http://www.swift.ac.uk/burst analyser/} (Evans et al. 2010).
The magnetar signature typically exhibits a shallow decay phase (or plateau) followed by a steeper
decay segment (a normal decay for canonical light curves or a very steep decay for internal plateaus)
when it is spinning down by losing rotational energy via MD and GW radiations.

Three criteria are adopted for our sample selection. First, we focus on those long GRBs that show such
a transition in the X-ray light curves, but required that the decay slope of the steeper decay segment
following the plateau phase should be in range of -1 to -2. Those are the typical decay slopes when GW
and MD radiations are dominant, respectively\footnote{The magnetar may be unstable and collapse into
the black hole before it spins down if the decay slope is steeper than 2.} (Zhang \& M\'esz\'aros 2001;
Lasky \& Glampedakis 2016; L\"{u} et al. 2018). Second, GRBs with bright X-ray flares\footnote{Bright
X-ray flares are defined as $F_{\rm p}/F_{\rm b}>5$, where $F_{\rm p}$ and $F_{\rm b}$ are the flux at
the peak time and the corresponding underlying flux, respectively (Hu et al. 2014; Zou et al. 2018).}
observed during the plateau or normal decay phase are excluded from our sample. Those flares are
believed to be from later activity of the central engine (Zhang et al. 2007; Peng et al. 2014). Third,
in order to estimate the intrinsic luminosity of the plateau emission, the redshift for our sample
needs to be measured. By adopting these criteria for our sample selection, 45 long GRBs are included in
our sample (also see Zou et al. 2018). Moreover, GRB 060614 is not included in our sample because its
properties satisfy the above criteria, a possible kilonova signature in the near-infrared band points
to a compact binary merger origin (Yang et al. 2015).

\subsection{Data Fitting with the Magnetar Model}
As a neutron star forms when a massive star collapses, its angular frequency evolves with time. The
solution of Eq.(\ref{torque equation}) can be expressed as
\begin{eqnarray}
\Omega (t)=\Omega_{0}(1+\frac{t}{\tau})^{\frac{1}{1-n}},
\label{Omega evolution}
\end{eqnarray}
where $\Omega_{0}$ is the initial angular frequency at $t=0$, and
$\tau=\frac{\Omega_{0}^{1-n}}{(n-1)k}$ is the spindown timescale of the magnetar. The observed
luminosity should be equivalent to $L_{\rm EM}$, namely
\begin{eqnarray}
L_{\rm obs}=L_{\rm EM}&=&\frac{B^2_{\rm p}R^{6}\Omega^{4}}{6c^{3}}\nonumber \\
&=&L_0(1+\frac{t}{\tau})^{\frac{4}{1-n}}
\label{L obs}
\end{eqnarray}
where $L_0=\frac{B^2_{\rm p}R^{6}\Omega_{0}^{4}}{6c^{3}}$ is the spindown luminosity of the magnetar.

\begin{figure*}
\centering
\begin{tabular}{cccccccccc}
\includegraphics[angle=0,scale=0.60]{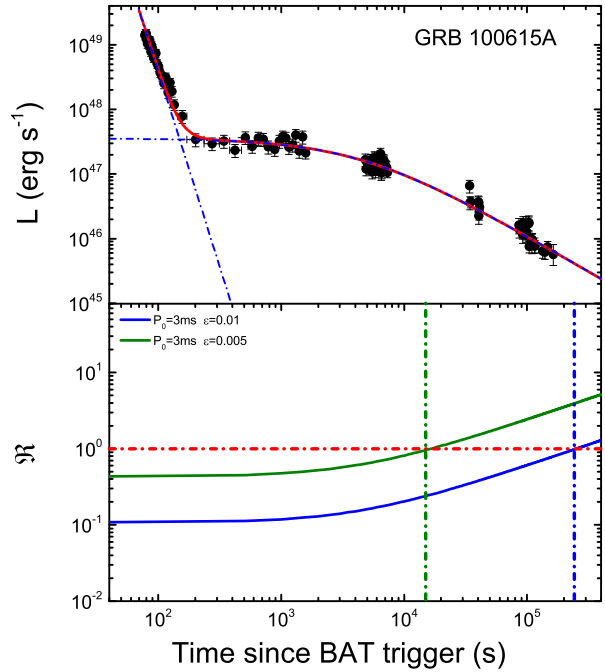}
\end{tabular}
\caption{An example light-curve fitting for GRB 100615A. Upper panel: the X-ray light curves with
power-law and magnetar models fits (blue dashed-dotted lines). The red solid line is the superposition
of power-law and magnetar models. Lower panel: the ratio between the EM luminosity and GW luminosity as
a function of time with $\varepsilon=0.01$ (blue solid line) and $\varepsilon=0.005$ (olive solid line)
by assuming $P_0=3$ ms.} \label{fig:XRTLC}
\end{figure*}

The observed plateau luminosity of the X-ray emission ($L_{\rm X}$), which is calculated in the
(0.3-10)keV energy band by assuming $10\%$ radiation efficiency, is equivalent to $L_{\rm obs}$. Thus,
Equation (\ref{L obs}) shows the plateau luminosity $L_{\rm X}\approx L_{0}$ for $t\ll \tau$, and a
power-law decay $L_{\rm X}\propto t^{\frac{4}{1-n}}$ for later times $t\gg \tau$. It is worth noting
that Eq. (\ref{L obs}) recovers the evolution of the luminosity with dominant MD radiation for $n=3$,
and the spindown timescale becomes the MD-dominated spindown timescale ($\tau=\tau_{\rm em}$). By the
same token, for $n=5$, the luminosity evolution is consistent with GW radiation dominating the magnetar
spindown, and $\tau=\tau_{\rm gw}$ in this limit (Zhang \& M\'esz\'aros 2001; Lasky \& Glampedakis
2016; L\"{u} et al. 2018; Sarin et al. 2018). Therefore, we adopt Eq. (\ref{L obs}) and combine it with
an initial power-law decay $L_{\rm X}=At^{\rm -\alpha}$ to fit the X-ray light curves of our sample. By
fitting the X-ray data, one can obtain the parameters of the magnetar model and the power-law component
(e.g., $A$, $\alpha$, $L_{0}$, $\tau$, and $n$). An example of the light-curve fitting is shown in
Fig.\ref{fig:XRTLC}, and other samples are shown in this link:
http://astro.gxu.edu.cn/info/1067/1530.htm. The fitting results are presented in Table 1.

\section{Results}
Our purpose is to find out the variable braking index of a magnetar in our sample and the energy
partition of a magnetar between GW and MD radiations by using {\em Swift}/XRT data of long GRBs.

\subsection{The braking index of magnetars}
The top panel of Fig. \ref{fig:XRTLC} shows the XRT data, together with our fit using Eq.\ref{L obs}
and an initial power law that fits the steep decay phase (dashed-dotted lines). The solid red curve
shows the superposition of the magnetar model and power-law fits.

\begin{figure*}
\centering
\includegraphics[angle=0,scale=0.6]{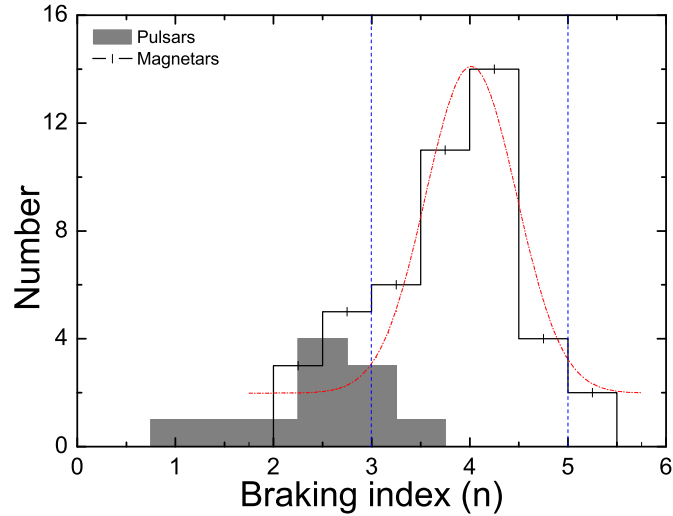}
\caption{Distribution of braking index (solid black line) and the best-fit Gaussian profile (red dashed
line). The gray filled histogram is the braking index of all known pulsar where the long-term spindown
is believed to be electromagnetically dominated. The two vertical dotted lines are corresponding to
$n=3$ and $n=5$, respectively.} \label{fig:Braking index}
\end{figure*}

Fig. \ref{fig:Braking index} shows the distribution of the braking indices of the magnetars in our
sample. Interestingly, we find that the braking indices are neither close to 3 nor 5, but range between
these two values. It is likely a normal distribution with the center value $n_{\rm c}=4.02\pm 0.11$. On
the other hand, we compare the braking index of our sample with that of all known pulsars in which the
long-term spindown is believed to be electromagnetically dominated (Antonopoulou et al. 2015; Archibald
et al. 2016; Clark et al. 2016). From the statistical point of view, the distribution of braking
indices of our sample is much larger than the distribution of pulsar braking indices, and it indicates
that the millisecond magnetar in our sample is intrinsically different from that in normal pulsar.
However, due to the limited data, this difference also seems to be caused by a selection effect. To
determine whether the distribution of braking indices of our sample is statistically consistent with
the distribution of pulsar braking indices, more observational data in the future is needed.

\subsection{The fraction of magnetar energy released via GW and MD radiations}
The braking index of a magnetar is reflected in the behavior of the energy released. With $n=3$, the
neutron star is spun down only via a dipole magnetic field in vacuum, while $n=5$ implies that the
neutron star is spun down through GW radiation (Lasky \& Glampedakis 2016; L\"{u} et al. 2018).
However, the fraction of energy released by the magnetar as MD and GW radiations when the value of $n$
in the range of 3-5 remains an open question. In this section, we investigate the evolution with time
of the fraction of energy released during magnetar spindown.

Here, we quantify the fraction of energy released by the magnetar by defining the ratio parameter
$\Re$,
\begin{eqnarray}
\Re &=&\frac{L_{\rm EM}}{L_{\rm GW}}=\frac{B^2_{\rm p}R^{6}\Omega^{4}}{6c^{3}}\cdot
\frac{5c^{5}}{32GI^{2}\epsilon^{2}\Omega^{6}}\nonumber \\
&=&\frac{125}{8192}\cdot \frac{L_0 c^5}{\pi^6 G M^2\epsilon^2R^4}\cdot P_{0,-3}^{6}\cdot
(1+\frac{t}{\tau})^{\frac{2}{n-1}}
\label{ratio}
\end{eqnarray}
where $L_0$, $\tau$, and the braking index $n$ are measured by fitting the X-ray data of long GRBs with
the magnetar model. The radius and mass of the neutron star depends on the equation of state. $\Re \gg
1$ implies that the spindown of the magnetar is dominated by MD radiation, while $\Re \ll 1$ implies
that the magnetar is spun down through GW radiation. The energy released during magnetar spindown is
contributed by both MD and GW radiations when $\Re$ is close to 1. The bottom panel of Fig.
\ref{fig:XRTLC} shows $\Re =\frac{L_{\rm EM}}{L_{\rm GW}}$ as a function of time by assuming $P_0=3$ ms
and $\epsilon=0.01$ and 0.005 with equation of state GM1 (Lasky et al. 2014; L\"{u} et al. 2018).

Given the initial parameters $P_0=3$ ms and $\epsilon=0.005$ and $0.01$, the distributions of $\Re$
tend to normal with center values $0.47\pm0.09$ and $0.12\pm0.04$, respectively. It suggests that a
large fraction of GRBs in our sample with energy released during magnetar spindown are dominated by GW
radiation for larger $\epsilon$ values. Such type of magnetar was also applied in some short GRBs to
constrain the ellipticity (Fan et al. 2013; Lasky \& Glampedakis 2016).

\begin{figure*}
\centering
\includegraphics[angle=0,scale=0.62]{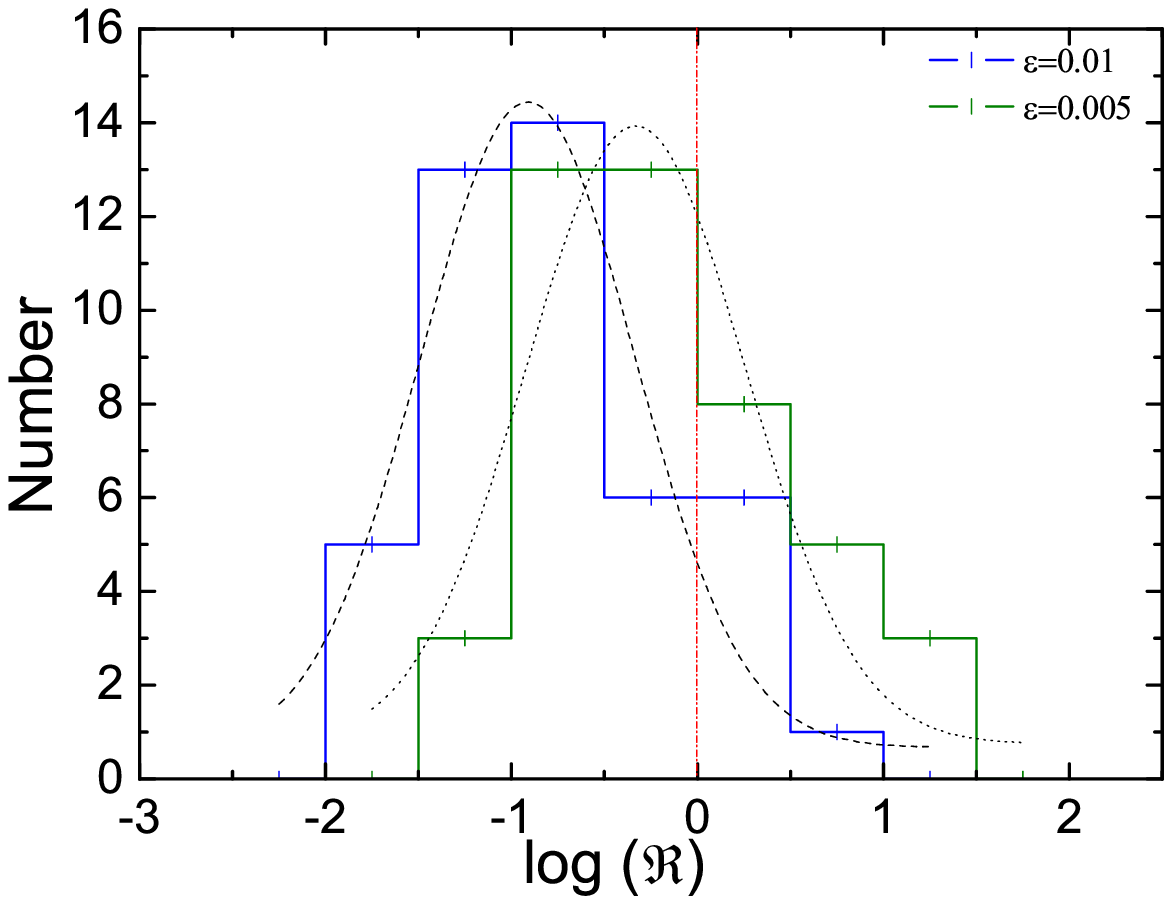}
\includegraphics[angle=0,scale=0.7]{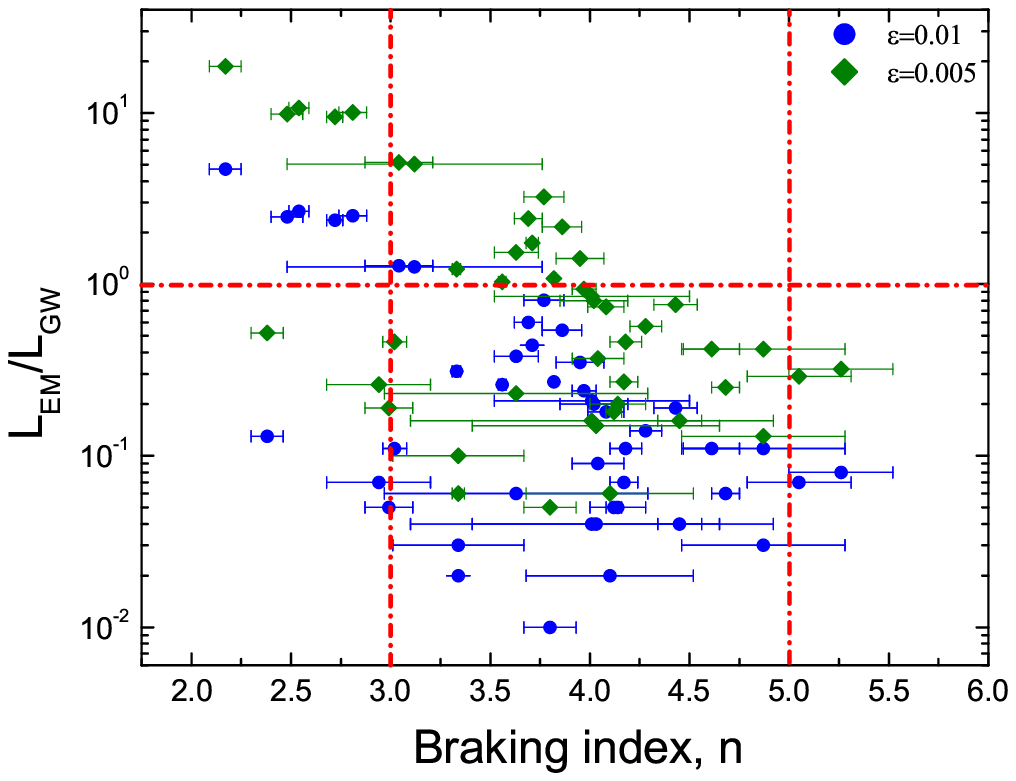}
\caption{Left: distributions of $\Re$ and its best-fit Gaussian profile for $\varepsilon=0.01$ (blue
solid line) and $\varepsilon=0.005$ (olive solid line). Right: derived the ration between luminosity of
EM and GW radiation as function of braking index with $\varepsilon=0.01$ (black solid points) and
$\varepsilon=0.005$ (blue solid diamond). The horizontal dash-dotted line is $\Re=1$, and two vertical
dash-dotted lines are $n=3$ and $n=5$, respectively.} \label{fig:ratio}
\end{figure*}

On the other hand, we investigate how the initial ratio parameter between $L_{\rm EM}$ and $L_{\rm
GW}$, $\Re$, is related to the braking index $n$. Figure \ref{fig:ratio} presents the $\Re-n$
correlation at $t=0$ for given $P_0=3$ ms by assuming $\epsilon=0.01$ and $\epsilon=0.005$,
respectively. There seems to be an anticorrelation between $\Re$ and $n$, namely a higher braking index
corresponds to a lower $\Re$ value within a large systematic error. If this anticorrelation indeed
exists, it indicates that a higher braking index tends to givenrise to GW-dominated radiation for
magnetar spindown, and that is consistent with the theoretical derivation in Zhang \& M\'esz\'aros
(2001). However, the derived $\Re$ value depends on the equation of state, initial period $P_0$, and
ellipticity $\epsilon$, which may be constrained by detecting GW radiation and its associated EM
emission simultaneously (Lasky et al. 2014; Li et al. 2016; L\"{u} et al. 2018).

\section{Conclusions and Discussion}
The observed X-ray plateau following a steep decay emission in the X-ray light curves of long GRBs is
believed to be from the ongoing energy injection of a magnetar and its spindown. In this paper, we
systematically analyzed the X-ray light curves of long GRBs detected by {\em Swift}/XRT before 2018
July. There are 45 long GRBs with redshifts measured whose X-ray light cueves exhibited a plateau
following a steep decay emission. By assuming that the central engines of these GRBs are newly born
magnetars, we measure the braking index $n$ of the putative millisecond magnetars via the observed
X-ray emission with MD and GW radiations. We define a dimensionless parameter $\Re$ that indicates the
fraction of energy released as MD and GW radiations and try to find out the distribution of $\Re$
initially and its relation with the braking index $n$. The following interesting results are obtained.
\begin{itemize}
 \item The inferred braking indices of the magnetars in our sample are neither close to 3 (i.e.,
     for magnetar spindown dominated by MD radiation) nor to 5 (i.e., for magnetar spindown
     dominated by GW radiation), but range between them. We find that the distribution of braking
     indices is likely normal with the center value $n_{\rm c}=4.02\pm 0.11$. That value is much
     larger than that for the distribution of known pulsars, where the long-term spindown is
     believed to be electromagnetically dominated. It indicated that the energy released by the
     millisecond magnetars in our sample is intrinsically different from those released by normal
     pulsars if this difference is not from a selection effect.
 \item The distribution of $\Re$ tends to normal with center values $0.47\pm0.09$ and $0.12\pm0.04$
     for $P_0=3$ ms and $\epsilon=0.005$ and 0.01, respectively. It indicates that for a small
     fraction of GRBs in our sample, the energy loss during magnetar spindown seems to be
     MD-dominated.
 \item The ratio parameter $\Re$ and braking index $n$ seem to be anticorrelated within a large
     systematic error at $t=0$, although it depends on the values of parameters $P_0$ and
     $\epsilon$. It suggest that a larger braking index leads to GW dominating the energy released
     during magnetar spindown.
\end{itemize}

The braking indices of most long GRBs in our sample range from 3 up to 5, but a small fraction of long
GRBs fall out of this range. Several possible models discussed in the literature are invoked to explain
the observed anomalous $n<3$ braking indices. For example, a twisted magnetosphere consisting of a
strong mixed poloidal-toroidal field increases the spindown torque in comparison to orthogonal vacuum
dipoles (Thompson et al. 2002; Kiuchi et al. 2011; Turolla et al. 2015), or a different magnetosphere
that is dipole force-/twist-free is used (Contopoulos \& Spitkovsky 2006). Another possible explanation
for $<3$ braking indices is related to the evolution of the included angle, which is the angle between
the spin axis and its surface dipole magnetic field axis (Lyne et al. 2013; 2015), or to the momentum
loss during a neutron star spindown due to both dipole radiation and a episodic or continuous particle
winds (Harding et al. 1999).

In comparison to the possible explanations for $n<3$ braking indices, there are fewer proposed models
for interpreting the $n>5$ braking indices. For example, if the magnetar spindown is through unstable
$r$ modes, then the braking index can be as large as 7 (Owen et al. 1998). However, the braking index
of a magnetar for unstable $r$ modes is also insensitive to both microscopic details and the saturation
amplitude. If this is the case, the true value of the braking index can be in range of 5-7 (Alford \&
Schwenzer 2014, 2015). On the other hand, the ratio parameter $\Re$ depends on the initial period and
ellipticity of the magnetar in our calculations, and the true value of $\Re$ can be constrained if the
initial period of the ellipticity can be measured by using independent methods in the future.

\acknowledgments We acknowledge the use of the public data from the {\em Swift} data archive and the UK
{\em Swift} Science Data Center. This work is supported by the National Basic Research Program (973
Programme) of China 2014CB845800, the National Natural Science Foundation of China (grant Nos.11603006,
11533003 and 11851304), Guangxi Science Foundation (grant Nos. 2017GXNSFFA198008, 2016GXNSFCB380005 and
2017AD22006), the One-Hundred-Talents Program of Guangxi colleges, the high level innovation team and
outstanding scholar program in Guangxi colleges, Scientific Research Foundation of Guangxi University
(grant no XGZ150299), and special funding for Guangxi distinguished professors (Bagui Yingcai \& Bagui
Xuezhe).



\begin{center}
\begin{deluxetable}{lllllllllllll}
\tablewidth{0pt} \tabletypesize{\footnotesize} \tabletypesize{\tiny} \tablecaption{The observational
properties and fitting results of our sample.}\tablenum{1}

\tablehead{\colhead{GRB}& \colhead{$z$}& \colhead{$T_{90}$}& \colhead{$L_{0}$\tablenotemark{a}}&
\colhead{$\tau$\tablenotemark{a}}& \colhead{$n$\tablenotemark{b}}&
\colhead{$\chi^{2}$/dof\tablenotemark{c}}&
\colhead{References\tablenotemark{d}}\\
\colhead{(Name)}&\colhead{Redshift}& \colhead{(second)}&\colhead{($10^{47}\rm erg~s^{-1}$)}&
\colhead{($10^3\rm~s$)}&\colhead{}& \colhead{}& \colhead{}}

\startdata
\hline 																									

050319	&	3.24	&	153 	&	6.43	$\pm$	0.17 	&	7.42 	$\pm$	0.73 	&	4.14	$\pm$	
0.14 	&	1.27 	&	(1)	\\
050822	&	1.434	&	103 	&	0.447	$\pm$	0.01 	&	13.27 	$\pm$	1.14 	&	4.45	$\pm$	
0.11 	&	1.29 	&	(2)	\\
050922B	&	4.9	&	151 	&	0.74	$\pm$	0.04 	&	198.43 	$\pm$	53.14 	&	2.94	$\pm$	0.26

&	1.05 	&	(3)	\\
051016B	&	0.9364	&	4 	&	0.132	$\pm$	0.0004 	&	13.89 	$\pm$	1.53 	&	3.8	$\pm$	0.13 	&

1.17 	&	(4)	\\
060604	&	2.68	&	80 	&	1.05	$\pm$	0.04 	&	10.61 	$\pm$	1.15 	&	4.04	$\pm$	0.13

&	1.32 	&	(5)	\\
060605	&	3.8	&	115 	&	11	$\pm$	0.27 	&	8.91 	$\pm$	0.78 	&	2.48	$\pm$	0.08 	&

0.92 	&	(6)	\\

060714	&	2.71	&	115 	&	6.06	$\pm$	0.21 	&	3.10 	$\pm$	0.31 	&	3.86	$\pm$	
0.10 	&	1.27 	&	(7)	\\
060729	&	0.54	&	115 	&	0.175	$\pm$	0.0002 	&	90.99 	$\pm$	2.77 	&	3.34	$\pm$	
0.03 	&	1.25 	&	(8)	\\
061121	&	1.314	&	81 	&	14.6	$\pm$	0.18 	&	2.79 	$\pm$	0.09 	&	3.71	$\pm$	0.03

&	1.00 	&	(9)	\\
070129	&	2.3384	&	461 	&	0.705	$\pm$	0.02 	&	20.45 	$\pm$	1.90 	&	3.95	$\pm$	
0.12 	&	1.19 	&	(10)	\\
070306	&	1.496	&	209 	&	1.47	$\pm$	0.03 	&	71.94 	$\pm$	7.01 	&	2.38	$\pm$	
0.08 	&	1.42 	&	(11)	\\
070328	&	2.0627	&	75 	&	218	$\pm$	2.01 	&	0.67 	$\pm$	0.02 	&	3.33	$\pm$	0.02 	&

1.35 	&	(12)	\\
070508	&	0.82	&	21 	&	32.5	$\pm$	0.25 	&	0.61 	$\pm$	0.01 	&	3.56	$\pm$	0.02

&	1.00 	&	(13)	\\
080430	&	0.767	&	14 	&	0.275	$\pm$	0.01 	&	5.81 	$\pm$	0.34 	&	4.68	$\pm$	0.07

&	1.00 	&	(14)	\\
081210	&	2.0631	&	146 	&	0.282	$\pm$	0.01 	&	52.06 	$\pm$	14.20 	&	3.34	$\pm$	
0.33 	&	1.11 	&	(12)	\\
090404	&	3.0	&	84 	&	2.03	$\pm$	0.05 	&	19.77 	$\pm$	1.82 	&	3.63	$\pm$	
0.11 	&	

1.39 	&	(15)	\\
090516	&	4.109	&	208 	&	11.2	$\pm$	0.23 	&	12.34 	$\pm$	0.89 	&	2.81	$\pm$	
0.07 	&	1.08 	&	(16)	\\
090529	&	2.625	&	69 	&	0.105	$\pm$	0.01 	&	49.45 	$\pm$	23.12 	&	4.03	$\pm$	0.62

&	1.24 	&	(17)	\\
090618	&	0.54	&	113 	&	9.05	$\pm$	0.05 	&	1.43 	$\pm$	0.02 	&	3.82	$\pm$	
0.01 	&	1.17 	&	(18)	\\
091018	&	0.971	&	4 	&	13.5	$\pm$	0.28 	&	0.32 	$\pm$	0.02 	&	4.12	$\pm$	0.04

&	1.29 	&	(19)	\\
091029	&	2.752	&	39 	&	2.07	$\pm$	0.04 	&	11.19 	$\pm$	0.78 	&	4.08	$\pm$	0.09

&	1.09 	&	(20)	\\
100302A	&	4.813	&	18 	&	0.678	$\pm$	0.03 	&	11.51 	$\pm$	2.08 	&	5.05	$\pm$	0.26

&	0.80 	&	(21)	\\
100615A	&	1.398	&	39 	&	3.56	$\pm$	0.09 	&	4.45 	$\pm$	0.48 	&	4.61	$\pm$	0.14

&	1.01 	&	(22)	\\
100814A	&	1.44	&	175 	&	1.28	$\pm$	0.02 	&	71.53 	$\pm$	3.96 	&	3.02	$\pm$	
0.06 	&	1.78 	&	(23)	\\
110808A	&	1.348	&	48 	&	(7.42	$\pm$	0.46)e-2 	&	13.46 	$\pm$	3.89 	&	4.87	$\pm$	
0.41 	&	1.08 	&	(24)	\\
111008A	&	4.9898	&	63 	&	25.8	$\pm$	0.54 	&	6.46 	$\pm$	0.42 	&	3.69	$\pm$	0.07

&	1.11 	&	(25)	\\
111228A	&	0.714	&	101 	&	0.751	$\pm$	0.01 	&	6.79 	$\pm$	0.37 	&	3.97	$\pm$	
0.06 	&	1.01 	&	(26)	\\
120422A	&	0.283	&	7 	&	(6.39	$\pm$	0.46)e-4 	&	165.35 	$\pm$	77.30 	&	3.63	$\pm$	
0.66 	&	1.12 	&	(27)	\\
120521C	&	6.0	&	27 	&	1.88	$\pm$	0.13 	&	11.83 	$\pm$	6.19 	&	3.12	$\pm$	
0.64 	&	

1.08 	&	(28)	\\
130609B	&	1.3	&	211 	&	26.5	$\pm$	0.32 	&	4.03 	$\pm$	0.17 	&	2.72	$\pm$	0.04

&	1.00 	&	(29)	\\
131105A	&	1.686	&	112 	&	2.25	$\pm$	0.07 	&	4.01 	$\pm$	0.59 	&	4.02	$\pm$	
0.17 	&	1.13 	&	(30)	\\
140430A	&	1.6	&	174 	&	0.563	$\pm$	0.02 	&	5.59 	$\pm$	1.35 	&	4.87	$\pm$	0.41

&	1.17 	&	(31)	\\
140703A	&	3.14	&	70 	&	20.9	$\pm$	0.57 	&	20.13 	$\pm$	2.05 	&	2.17	$\pm$	0.08

&	1.19 	&	(32)	\\
160227A	&	2.38	&	317 	&	3.04	$\pm$	0.08 	&	23.52 	$\pm$	1.99 	&	3.77	$\pm$	
0.10 	&	1.49 	&	(33)	\\
160804A	&	0.736	&	144 	&	0.118	$\pm$	0.0004 	&	6.32 	$\pm$	0.99 	&	5.26	$\pm$	
0.26 	&	1.08 	&	(34)	\\
161117A	&	1.549	&	126 	&	3.84	$\pm$	0.08 	&	4.15 	$\pm$	0.26 	&	4.18	$\pm$	
0.08 	&	1.12 	&	(35)	\\
170113A	&	1.968	&	21 	&	16.4	$\pm$	0.32 	&	1.56 	$\pm$	0.10 	&	4.17	$\pm$	0.07

&	1.24 	&	(36)	\\
170519A	&	0.8181	&	216 	&	0.524	$\pm$	0.01 	&	11.70 	$\pm$	1.20 	&	2.99	$\pm$	
0.12 	&	1.24 	&	(37)	\\
170531B	&	2.366	&	164 	&	0.952	$\pm$	0.07 	&	4.88 	$\pm$	1.61 	&	4.01	$\pm$	
0.49 	&	1.20 	&	(38)	\\
170607A	&	0.557	&	320 	&	0.187	$\pm$	0.0004 	&	16.20 	$\pm$	1.26 	&	4.43	$\pm$	
0.11 	&	1.45 	&	(39)	\\
170705A	&	2.01	&	207 	&	6.15	$\pm$	0.13 	&	9.47 	$\pm$	0.59 	&	4.28	$\pm$	
0.08 	&	1.77 	&	(40)	\\
171222A	&	2.409	&	175 	&	0.114	$\pm$	0.01 	&	126.70 	$\pm$	71.35 	&	4.01	$\pm$	
0.91 	&	1.36 	&	(41)	\\
180325A	&	2.25	&	94 	&	114	$\pm$	2.23 	&	2.99 	$\pm$	0.20 	&	2.54	$\pm$	0.05 	&

1.01 	&	(42)	\\
180329B	&	1.998	&	210 	&	2.57	$\pm$	0.08 	&	8.17 	$\pm$	1.16 	&	3.04	$\pm$	
0.17 	&	1.30 	&	(43)	\\
180404A	&	1.0	&	701 	&	0.182	$\pm$	0.01 	&	7.73 	$\pm$	3.11 	&	4.1	$\pm$	
0.42 	&	

1.02 	&	(44)	\\

\enddata
\tablenotetext{a}{The plateau luminosity and break time of the X-ray light curves in our sample.}
\tablenotetext{b}{The derived braking index of magnetar.} \tablenotetext{c}{The goodness of the X-ray
light-curve fitting using the magnetar model and power-law component if possible.}
\tablenotetext{d}{The references for the redshift in our sample.}

\tablerefs{(1)Fynbo et al. 2005;(2)Hjorth et al. 2012;(3)Perley et al. 2016;(4)Soderberg et al.
2005;(5)Castro-Tirado et al. 2006;(6)Peterson \& Schmidt 2006;(7)Jakobsson et al. 2006;(8)Thoene et al.
2006;(9)Bloom et al. 2006;(10)Kruehler et al. 2012;(11)Jaunsen et al. 2007;(12)Kruehler et al.
2015;(13)Jakobsson et al. 2007;(14)Cucchiara \& Fox. 2008;(15)Perley et al. 2013;(16)de Ugarte Postigo
et al. 2009;(17)Malesani et al. 2009;(18)Cenko et al. 2009;(19)Chen et al. 2009;(20)Chornock et al.
2009;(21)Chornock et al. 2010;(22)Kruehler et al. 2013;(23)O'Meara et al. 2010;(24)de Ugarte Postigo et
al. 2011;(25)Wiersema et al. 2011;(26)Dittmann et al. 2011;(27)Schulze et al. 2012;(28)Tanvir et al.
2012;(29)Ruffini et al. 2013;(30)Xu et al. 2013;(31)Kruehler et al. 2014;(32)Castro-Tirado et al.
2014;(33)Xu et al. 2016a;(34)Xu et al. 2016b;(35)Malesani et al. 2016;(36)Xu et al. 2017;(37)GCN
21119;(38)GCN 21177;(39)GCN 21240;(40)de Ugarte Postigo et al. 2017a; (41)de Ugarte Postigo et al.
2017b;(42)Heintz et al. 2018;(43)Izzo et al. 2018;(44)Selsing et al. 2018}

\end{deluxetable}
\end{center}



\end{document}